\documentclass[manuscript,screen]{acmart}

\acmJournal{TOSEM}
\setcopyright{acmlicensed}
\copyrightyear{2026}
\acmYear{2026}
\acmDOI{XXXXXXX.XXXXXXX}
\acmVolume{0}
\acmNumber{0}
\acmArticle{0}
\acmMonth{0}

\usepackage{amsmath}
\usepackage{amsfonts}
\usepackage{graphicx}
\usepackage{textcomp}
\usepackage{url}
\usepackage{booktabs}
\usepackage{tabularx}
\usepackage{xspace}
\usepackage{stfloats}
\usepackage[normalem]{ulem}
\usepackage{changepage}
\usepackage[normalsize]{subfigure}
\usepackage{multirow}
\usepackage{paralist}
\usepackage{balance}
\usepackage{breqn}
\usepackage{makecell}
\usepackage{enumitem}
\usepackage{multicol}
\usepackage{bm}
\usepackage[ruled,linesnumbered]{algorithm2e}
\usepackage{colortbl}
\usepackage{listings}
\usepackage{xcolor}  

\usepackage{tikz}
\usetikzlibrary{positioning,fit,backgrounds,arrows.meta,calc}

\usepackage{pifont}
\newcommand{\cmark}{\ding{51}}
\newcommand{\xmark}{\ding{55}}

\acmJournal{TOSEM}
\begin{document}

\title{Formal-Method-Guided Vibe Coding: Closing the Verification Loop on AI-Generated Safety-Critical Software Through Model-Driven Engineering}


\author{Ran Wei}
\email{r.wei5@lancaster.ac.uk}
\orcid{0000-0003-2191-1359}
\affiliation{%
  \institution{Lancaster University}
  \city{Lancaster}
  \country{United Kingdom}
}

\author{Le Zhu}
\authornote{Corresponding Author}
\email{l-zhu19@mails.tsinghua.edu.cn}
\orcid{0009-0009-8762-0619}
\affiliation{%
  \institution{Tsinghua University}
  \city{Beijing}
  \country{China}
}

\author{Haochi Wang}
\orcid{0009-0005-3768-5335}
\email{wanghaochi1222@gmail.com}
\affiliation{%
  \institution{Harbin Institute of Technology}
  \city{Harbin}
  \country{China}}

\author{Jim Woodcock}
\email{jim.woodcock@york.ac.uk}
\orcid{0000-0001-7955-2702}
\affiliation{%
  \institution{Southwest University, China; Aarhus University, Denmark; University of York}
  \city{York}
  \country{UK}}

\author{Fang Yan}
\email{fang.yan@york.ac.uk}
\orcid{0000-0001-5603-3467}
\affiliation{%
  \institution{University of York}
  \city{York}
  \country{UK}}

\author{Simon Foster}
\email{simon.foster@york.ac.uk}
\orcid{0000-0002-9889-9514}
\affiliation{%
  \institution{University of York}
  \city{York}
  \country{UK}}
  
\author{Xiangyang Ji}
\email{xyji@tsinghua.edu.cn}
\orcid{0000-0001-9542-5260}
\affiliation{%
  \institution{Tsinghua University}
  \city{Beijing}
  \country{China}
}


\begin{abstract}
Vibe coding---accepting LLM-generated source from natural-language intent with minimal review---is fast and may be adequate for low-criticality consumer software.
But for safety-critical systems governed by DO-178C, IEC~61508, or ISO~26262, it offers no path to certification: large language models (LLMs) provide no formal correctness guarantees~\cite{Sun2024Clover}, and existing remedies target verification-aware languages (Dafny, Verus, Lean) that are scarce in pretraining data and absent from industrial toolchains.
\par
This paper closes the gap.
We present \textsc{Forge} (\textbf{F}ormal method \textbf{O}riented \textbf{R}efinement loop for \textbf{GE}nerated code): a closed-loop pipeline that \emph{guides vibe coding through formal verification} using established Model-Driven Engineering (MDE) infrastructure.
Through vibe coding, we generate Java source code; our pipeline then extracts---via model transformations---formal artefacts in three different formalisms, each checked by a complementary verifier: deductive verification (Dafny), Communicating Sequential Processes (CSP) refinement via the Failures-Divergences Refinement checker (FDR4), and theorem proving using \emph{Z-Machines}\footnote{A Z-style state-machine notation within the Isabelle/HOL ecosystem~\cite{Yan2023CompositionalVerification}.} in Isabelle; every verification failure becomes a structured correction prompt that drives the next code-generation iteration.
The LLM is the draft generator, the MDE chain is the discriminator, and the developer never has to read the formal models.
\par
Empirically, we find that the pipeline produces standards-relevant verification evidence for LLM-generated Java---a step toward certification.
\end{abstract}



\maketitle




\section{Introduction}
\label{sec:introduction}


Vibe coding has reshaped how software is developed~\cite{Fawzy2025VibeCoding, Fan2023LLM4SE, Pimenova2025goodvibrations}: under the banner coined by Karpathy~\cite{Karpathy2025}, developers describe intent in natural language to a large language model (LLM), accept the generated source with minimal review, and iterate through dialogue rather than manual editing.
For consumer software, the outputs are good enough, but the speed of adoption has outpaced assurance: LLM-generated code can pass standard benchmarks yet fail stronger tests, and carries no formal correctness guarantees~\cite{Sun2024Clover} (we return to this evidence in Section~\ref{sec:background:vibecoding}).
For safety-critical domains governed by certification standards such as DO-178C~\cite{DO178C2011}, IEC~61508~\cite{IEC61508}, and ISO~26262~\cite{ISO26262}, where traceable evidence of behavioural correctness is mandatory, unguided vibe coding is unsuitable in its present form.


The research community has responded along two principal lines, but both remain disconnected from mainstream industrial practice.
The \textbf{first line} targets verification-aware intermediate languages such as Dafny~\cite{Leino2010Dafny}, Verus~\cite{Lattuada2024Verus}, and Lean~\cite{Moura2021Lean4}. 
These approaches direct LLMs to generate code in languages whose type systems and built-in annotations support automated proof. 
Results are improving, with recent work showing that LLMs can synthesize verified Dafny methods for a nontrivial fraction of benchmark tasks~\cite{Misu2024Dafny}, but three structural barriers limit industrial adoption.
First, these languages are scarce in LLM pretraining data, limiting generation quality.
Second, the DAFNYCOMP benchmark~\cite{DAFNYCOMP2025} shows that verification success drops by 92\% when compositional reasoning across function boundaries is required, and the CLEVER benchmark~\cite{CLEVER2025} finds that current LLMs solve approximately 1 of 161 end-to-end verified problems in Lean.
Third, and most relevant to practice, verification-aware languages are largely absent from industrial software toolchains: organisations building safety-critical systems in Java, C, or C++ cannot readily adopt Dafny or Lean without abandoning their existing modelling, testing, and certification infrastructure.
More broadly, formal methods have a steep learning curve that has historically limited their uptake beyond specialist teams~\cite{Woodcock2009FormalMethods}, making any approach that requires engineers to write or reason about formal specifications a bottleneck to adoption.
The \textbf{second line} adds \emph{post-hoc verification layers}---Clover~\cite{Sun2024Clover} triangulates code, annotations, and docstrings; SYNVER~\cite{Mukherjee2025SYNVER} pairs generated C programs with Rocq proofs---but these treat verification as a filter rather than a feedback signal, offering no mechanism to guide the generator toward correctness.


In this paper, we take a different path with \textsc{Forge} (\textbf{F}ormal method \textbf{O}riented \textbf{R}efinement loop for \textbf{GE}nerated code).
Rather than targeting niche verification languages or treating verification as a post-hoc filter, \textsc{Forge} routes the generated code through \emph{established Model-Driven Engineering (MDE) infrastructure}---the modelling, transformation, and code-generation toolchains already deployed in industrial software and systems engineering---and uses formal verification failures as iterative correction prompts that guide subsequent generation passes.
The key observation is that mainstream languages like Java, which dominate both LLM training corpora and industrial practice, can be systematically transformed into formal models amenable to machine-checked verification, provided that \textit{the right model extraction and transformation pipeline is in place}.
This approach sidesteps the language adoption barrier entirely: developers and LLMs continue to work in Java, while verification is performed on extracted formal models that developers never need to see.
Verification applies to Java within an \emph{extraction-tractable profile}---a restricted subset in the long tradition of safety-critical language profiles (MISRA, Ravenscar, Safety-Critical Java), but motivated here by faithful source-to-model extraction rather than the runtime predictability those profiles target (Section~\ref{sec:discussion:javasubsets}).


We present a pipeline organised into three families: \emph{requirements and generation} (Phases 1--2)---Phase 1 performs requirements engineering to produce the requirement set, and Phase 2 generates Java via a five-layer top-down prompt sequence; \emph{extraction} (Phases 3--5) parses the Java into a Spoon EMF model, transforms it into a RoboChart~\cite{Miyazawa2019RoboChart} state machine, and emits Dafny specifications, CSP-M scripts, and Z-Machine theories in Isabelle; \emph{verification with feedback} (Phases 6--7) discharges those artefacts via Dafny, FDR4, and Isabelle, and translates failures into structured fix directives.
We demonstrate the pipeline on three case studies of increasing complexity: SRanger, a small single-controller ground-robot drawn from the published RoboStar~\cite{Miyazawa2019RoboChart} case-study set\footnote{\url{https://robostar.cs.york.ac.uk/case_studies/sranger/index.html}} whose external authorship controls for the ``cherry-picked case studies'' concern; the Last Response Engine (LRE)~\cite{Foster2020AUV,Wei2024ACCESS}, a single-controller four-mode AUV safety governor with an independent published ground truth; and another case from the RoboStar case-study set: a Chemical Detector autonomous mobile robot\footnote{\url{https://robostar.cs.york.ac.uk/case_studies/autonomous-chemical-detector/autonomous-chemical-detector.html}}. 
The pipeline runs end-to-end on all three with no case-study-specific changes; across five independent runs per study (15 in total), every run converges to fully verified code. Two controls confirm that the verification feedback is doing the work: no cold, single-pass generation converges on its own (0 of 30 runs), and a compile-only ablation that discounts only the verifier feedback leaves every run verification-failing---so the verifier feedback specifically, not iteration alone, drives convergence.


The contributions of this paper are:

\begin{enumerate}
    \item \textbf{A closed draft-and-discriminate loop for verifiable code generation---our central contribution.} We show that LLM-generated mainstream code can be made formally verifiable neither by training the model on a verification-aware language nor by filtering its output once, but by routing it through established MDE infrastructure---EMF~\cite{Steinberg2008EMF}, Spoon~\cite{Pawlak2015Spoon}, Epsilon~\cite{Kolovos2008Epsilon}, RoboChart~\cite{Miyazawa2019RoboChart}---used as the \emph{discriminator}: verification failures become the next generation's correction signal, and the developer never reads a formal model.

    \item \textbf{The first EMF-native text-to-model transformation for modern Java \emph{(enabling contribution)}.} We contribute a Spoon-to-EMF bridge---an Ecore Java metamodel derived from Spoon's metamodel API together with a reflective discoverer that emits conforming EMF models---that recovers models from Java~17+ source (sealed interfaces, records, pattern syntax); no prior tool combines EMF-native models, modern Java, and active maintenance.

    \item \textbf{A complementary three-verifier design with a defect-level account \emph{(enabling contribution)}.} A single extracted Java model drives Dafny, FDR4, and Isabelle/HOL, and we document which class of defect each verifier catches and which it misses (Table~\ref{tab:complementarity}), establishing that the three are not interchangeable.

    \item \textbf{A fully traceable chain from requirement to proof.} Per-phase trace files link every natural-language requirement through the generated source, the Spoon EMF and RoboChart models, and the three formal artefacts, and are consumed by the feedback parsers to emit per-requirement fix directives.

    \item \textbf{Empirical evidence that the loop converges, and that the verifier feedback drives it.} On three external case studies, the unmodified pipeline converges to fully verified code on all 15 runs, while no single-pass cold generation does (0 of 30); an ablation attributes this to the verifier feedback specifically, not iteration alone.

    \item \textbf{An open-source implementation} of the pipeline, the three case studies, and all experiments.
\end{enumerate}


The paper is organised as follows.
Section~\ref{sec:background} covers background---vibe coding, model-driven engineering, and formal verification---and introduces the three case studies.
Section~\ref{sec:approach} presents the seven-phase pipeline architecture and its closed-loop refinement.
Section~\ref{sec:evaluation} demonstrates the pipeline end-to-end and answers five research questions, including a cold-baseline experiment and a compile-only ablation.
Section~\ref{sec:discussion} discusses implications, design rationale, generalisability, threats to validity, and reproducibility.
Section~\ref{sec:agenda} articulates a research agenda for the wider community.
Section~\ref{sec:conclusion} concludes.
\section{Background}
\label{sec:background}

This section introduces the three technical pillars on which our pipeline rests---vibe coding as a rising development paradigm, MDE as a model-centric development methodology, and formal verification as a correctness discipline---and then presents the three case studies used throughout the rest of the paper.

\subsection{Vibe Coding and LLM-Assisted Code Generation}
\label{sec:background:vibecoding}

\emph{Vibe coding}, coined by Karpathy~\cite{Karpathy2025}, describes a workflow in which a developer specifies intent in natural language, an LLM generates source code, and the developer accepts the output with minimal or no manual review.
Unlike traditional AI-assisted coding or pair programming, the developer may avoid close examination of the generated artefacts, relying instead on execution results and iterative dialogue with the model.

Adoption has been rapid: Y~Combinator reported that 25\% of startups in its Winter 2025 batch had codebases that were 95\% AI-generated~\cite{Fawzy2025VibeCoding}.
However, this speed comes at the cost of assurance: LLM-generated programs can pass standard benchmarks yet fail stronger tests~\cite{Liu2024EvalPlus}, exhibit recurring bug patterns~\cite{Tambon2025BugsLLM}, and lack formal correctness guarantees in ordinary generation workflows~\cite{Sun2024Clover}.

For safety-critical domains governed by standards such as DO-178C~\cite{DO178C2011}, IEC~61508~\cite{IEC61508}, and ISO~26262~\cite{ISO26262}, where certification demands traceable evidence of behavioural correctness, unguided vibe coding is therefore unsuitable in its present form.

The central tension motivating this paper is thus: vibe coding dramatically lowers the barrier to code generation, but the resulting artefacts carry no certification-grade evidence.
Our pipeline addresses this by treating LLM-generated code not as a final deliverable but as the input to a model-driven transformation and verification chain.

\subsection{Model-Driven Engineering}
\label{sec:background:mde}

Model-Driven Engineering (MDE) treats models as first-class engineering artefacts from which code, documentation, and analysis results are derived through automated model management operations~\cite{Kolovos2008Epsilon}.
A \emph{metamodel} defines a modelling language's abstract syntax; the Eclipse Modelling Framework (EMF)~\cite{Steinberg2008EMF} is the dominant Java-ecosystem implementation, with metamodels specified in Ecore and instances serialised in XMI~\cite{OMG2015XMI}.

Three kinds of transformation matter to our pipeline.
\emph{Text-to-model} (T2M) parses program text into model instances~\cite{durelli2014mapping}; MoDisco~\cite{Bruneliere2014MoDisco} is the established Java T2M tool, but its metamodel has remained frozen at JDK~5.
\emph{Model-to-model} (M2M) transforms a source-metamodel instance into a target-metamodel instance via transformation rules~\cite{Czarnecki2006MTSurvey,Sendall2003HeartSoul,Mens2006Taxonomy}; we use it to lift Java AST models to RoboChart state machines (Section~\ref{sec:approach:extraction}).
\emph{Model-to-text} (M2T) transforms models to textual artefacts via templates~\cite{Czarnecki2006MTSurvey}; established tools include the Epsilon Generation Language (EGL)~\cite{Rose2008EGL} and Acceleo~\cite{Acceleo2024}.
We generate three M2T targets from the extracted models: Dafny specifications directly from the Java model, and---from the RoboChart model derived from it by M2M---RoboChart textual models consumed by the official CSP generator and Z-Machine theories for Isabelle/HOL (Section~\ref{sec:approach:extraction}).

\subsection{Formal Verification of Safety-Critical Systems}
\label{sec:background:formal}

Certification standards govern safety-critical systems in aerospace, automotive, and industrial automation ---DO-178C~\cite{DO178C2011} (avionics), IEC~61508~\cite{IEC61508} (industrial), and ISO~26262~\cite{ISO26262} (automotive)---that classify software by criticality level (DAL / SIL / ASIL respectively) and require structured, traceable evidence that the implementation satisfies its safety requirements; formal methods are recognised as a verification supplement at the highest criticality levels.

This evidence is typically organised in an \emph{assurance case}---a structured argument, supported by evidence, that a system satisfies a defined set of (safety and/or security) claims~\cite{Kelly2004GSN,Wei2019SACM}.
Amongst the forms of evidence used to substantiate such cases, formal verification plays a particularly important role: it can provide mathematical guarantees that a system satisfies its specified properties---absence of deadlocks, adherence to safety invariants, refinement against an abstract specification---rather than relying solely on testing~\cite{Woodcock2009FormalMethods,Luckcuck2019FormalAutonomous}.
However, formal methods have historically seen limited industrial uptake, largely because they require specialised mathematical expertise that conventional software engineers rarely possess~\cite{Woodcock2009FormalMethods}.
Our pipeline addresses this by encapsulating formal verification behind automated model transformations, so that developers work exclusively in Java and need not interact with the formal models directly.

\textbf{Dafny}~\cite{Leino2010Dafny} is a verification-aware programming language developed by Microsoft Research that supports design-by-contract specification through preconditions, postconditions, loop invariants, and algebraic datatypes.
Programs are verified statically by translation to the Boogie intermediate verification language~\cite{Barnett2006Boogie} and discharge of proof obligations via the Z3 SMT solver.
Dafny's verification is \emph{deductive}: it reasons about individual functions and methods against their contracts, proving properties such as the absence of runtime errors, adherence to data invariants, and correct functional behaviour with respect to specifications.

\textbf{RoboChart}~\cite{Miyazawa2019RoboChart} is a domain-specific modelling language for robotic systems developed as part of the RoboStar framework.\footnote{\url{https://robostar.cs.york.ac.uk/}}
RoboChart semantics is based on CSP, with timed behaviour captured using tock-CSP~\cite{Baxter2022TockCSP}, which is a timed variant of the process algebra CSP that models both event-driven behaviour and the passage of discrete time.
RoboTool, the accompanying Eclipse-based tool, provides graphical and textual editors for RoboChart models and automatically generates CSP semantics from them.
The generated CSP is then verified by the FDR refinement checker~\cite{Gibson2014FDR}, which can establish deadlock freedom, divergence freedom, and failures-divergences refinement against a specification process.
RoboChart state machines serve as the intermediate formalism in our pipeline's M2M phase: they are sufficiently expressive to capture the reactive behaviour of safety controllers while retaining a formal semantics that enables automated property checking.

\textbf{Isabelle/HOL}~\cite{Nipkow2002Isabelle} is a generic interactive theorem prover.
The Z-Machine library~\cite{Yan2023CompositionalVerification} provides a Z-flavoured state-machine notation within the Isabelle/HOL ecosystem, together with built-in tactics for proving deadlock freedom and structural invariants of state transitions.
Z-Machines are particularly well suited to controllers expressed as Z-style \emph{zstores} (record-like state) with named \emph{zoperations} (guarded pre/update pairs); proofs of deadlock freedom are largely automatic, requiring user intervention only when the residual disjunction over states cannot be closed by enumeration.

The three verifiers are complementary: Dafny reasons over individual transitions and their contracts (function-level correctness), FDR4 reasons over the entire process algebra (behavioural properties of the full state machine), and Isabelle/HOL discharges deadlock freedom and zstore invariants over the symbolic transition relation (model-level structural properties).
Our pipeline exploits all three by generating Dafny, CSP, and Isabelle theories from the same extracted Java model.

\subsection{Case Studies}
\label{sec:background:casestudies}

We evaluate the pipeline on three case studies of increasing complexity: a single-controller ground robot with a timed transition, a single-controller safety governor with a four-mode guard hierarchy, and a multi-controller system whose controllers coordinate through 
events.
All three originate from published external sources rather than systems we designed for this paper---addressing the concern that authors might pick case studies that favour their tool.
For each study, the structured requirements specification, the natural-language system description, and the supporting prompt and annotation assets that drive the pipeline were authored as part of this work: a non-trivial requirements-engineering effort, because the available prior material is either loose textbook prose or earlier-formalism modelling work that does not match the schema the pipeline consumes.\footnote{The case-study assets are available in the \textsc{Forge} repository: \url{https://github.com/wrwei/Forge/tree/main/forge.assets/case-studies}.}

\textbf{SRanger} is a small autonomous ground robot taken from the published University of York RoboStar case-study set.\footnote{\url{https://robostar.cs.york.ac.uk/case_studies/sranger/index.html}}\footnote{\url{https://github.com/wrwei/Forge/tree/main/forge.assets/case-studies/sranger}} 
A single controller drives the vehicle forward, rotates in place when an obstacle is detected within a threshold distance, and terminates on an end-task command, across \emph{three} operating modes (Moving, Turning, Final) and 7 transitions.
Its requirements specification contains 23 individually-identified requirements across nine type categories.
The generated Java controller is approximately 220 lines across 10 files.

\textbf{The Last Response Engine} (LRE) is a reactive safety controller for an Autonomous Underwater Vehicle (AUV) developed by the National Oceanography Centre~\cite{Foster2020AUV} and previously modelled in RoboChart and verified in Isabelle/SACM~\cite{Wei2024ACCESS}.\footnote{\url{https://github.com/wrwei/Forge/tree/main/forge.assets/case-studies/lre}}
The LRE has \emph{one} controller class, \emph{four} operating modes---Operator Control Mode (OCM), Main Operating Mode (MOM), High Caution Mode (HCM), and Collision Avoidance Mode (CAM)---and 13 transitions.
Its requirements specification, included in the pipeline's case-study assets, contains 51 individually-identified requirements across 12 type categories: architectural, data type, constants, event, actuator, distance, operation, variable, state, guard predicate, transition, and constraint.
The generated Java controller is 613 lines across 12 files, organised into sub-packages for actuator, controller, event, mode, operation, sensor, constants, and annotations.

\textbf{The Chemical Detector} is an autonomous mobile robot whose mission is to locate the source of a target chemical (e.g.\ a gas leak) in an unknown environment that may contain obstacles.\footnote{\url{https://github.com/wrwei/Forge/tree/main/forge.assets/case-studies/chemical_detector}}
Unlike the LRE, the Chemical Detector's behaviour is the parallel composition of \emph{two} controllers: \texttt{GasAnalysisController} (five modes: Reading, Analysis, NoGas, GasDetected, Final) classifies gas-sensor readings, while \texttt{MovementController} (eight modes including obstacle-avoidance recovery) controls the robot's motion. 
The two are composed in parallel and coordinate through 
\texttt{turn}/\texttt{stop}/\texttt{resume} events (emitted by the Gas Analysis machine and consumed as triggers by the Movement machine), with no separate dispatcher.
The requirements specification contains 81 requirements across 11 type categories.
The generated Java code is 618 lines across the two controllers plus shared infrastructure.

The three case studies exercise distinct features of the pipeline.
SRanger stresses the M2M's clock-pattern classifier through a timed predicate; LRE adds a four-mode safety hierarchy of priority-ordered autonomous collision-avoidance transitions, gated on closest-distance- and time-to-closest-approach kinematics, that override operator commands; and the Chemical Detector adds multi-controller composition, typed event payloads (the \texttt{Gas} event carrying a \texttt{List<GasSensor>} of \texttt{Chem}-intensity records), and a terminal mode (\texttt{Final}) whose handling forced an explicit deadlock-freedom design rule.
Section~\ref{sec:evaluation} reports the end-to-end pipeline results for all three.


\section{Approach}
\label{sec:approach}

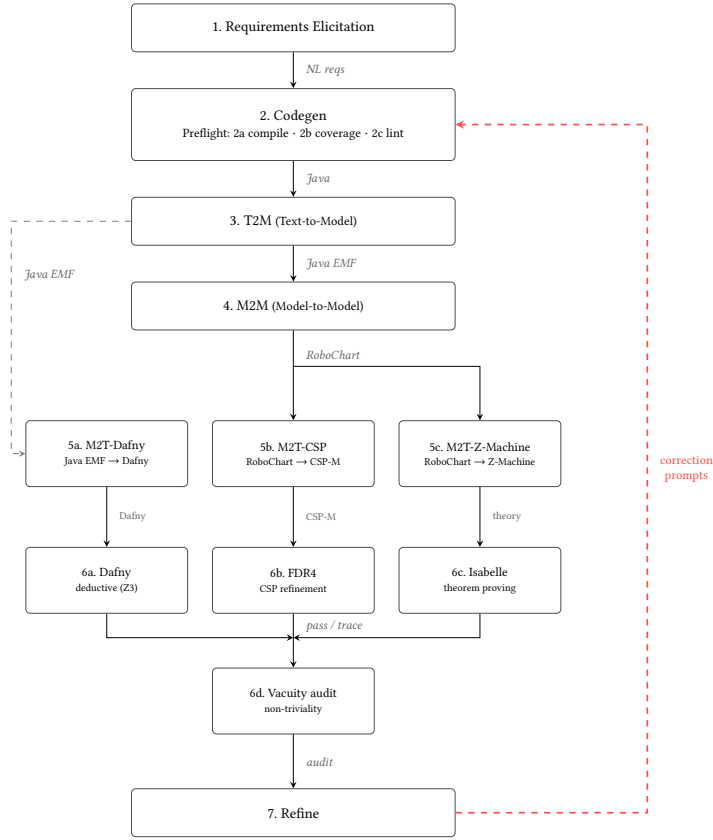
\begin{figure}[t]
\centering
\resizebox{0.62\textwidth}{!}{%
\begin{tikzpicture}[
    font=\footnotesize,
    >=stealth,
    node distance=6mm,
    block/.style={rectangle, draw=black!70, rounded corners=2pt, minimum width=54mm, minimum height=8mm, align=center, inner sep=2pt, fill=white},
    codeblock/.style={rectangle, draw=black!70, rounded corners=2pt, minimum width=54mm, minimum height=12mm, align=center, inner sep=2pt, fill=white},
    leaf/.style={rectangle, draw=black!70, rounded corners=2pt, minimum width=27mm, minimum height=11mm, align=center, inner sep=2pt, fill=white, font=\scriptsize},
    art/.style={font=\scriptsize\itshape, text=black!60},
    feedback/.style={->, dashed, draw=red!70, thick},
    bypass/.style={->, dashed, draw=black!50, thin}
]

\node[block] (req) {1.~Requirements Elicitation};
\node[codeblock, below=of req] (code) {2.~Codegen \\ \scriptsize Preflight: 2a compile $\cdot$ 2b coverage $\cdot$ 2c lint};
\node[block, below=of code] (t2m) {3.~T2M \scriptsize(Text-to-Model)};
\node[block, below=of t2m] (m2m) {4.~M2M \scriptsize(Model-to-Model)};

\node[leaf, below=15mm of m2m]                  (m2t_b) {5b. M2T-CSP \\ \tiny RoboChart $\to$ CSP-M};
\node[leaf, left=4mm of m2t_b]                  (m2t_a) {5a. M2T-Dafny \\ \tiny Java EMF $\to$ Dafny};
\node[leaf, right=4mm of m2t_b]                 (m2t_c) {5c. M2T-Z-Machine \\ \tiny RoboChart $\to$ Z-Machine};

\node[leaf, below=10mm of m2t_a] (v_a) {6a. Dafny \\ \tiny deductive (Z3)};
\node[leaf, below=10mm of m2t_b] (v_b) {6b. FDR4 \\ \tiny CSP refinement};
\node[leaf, below=10mm of m2t_c] (v_c) {6c. Isabelle \\ \tiny theorem proving};

\node[leaf, below=9mm of v_b] (vac) {6d. Vacuity audit \\ \tiny non-triviality};
\node[block, below=9mm of vac] (ref) {7.~Refine};

\draw[->] (req) -- node[art, right=1mm] {NL reqs} (code);
\draw[->] (code) -- node[art, right=1mm] {Java} (t2m);
\draw[->] (t2m) -- node[art, right=1mm] {Java EMF} (m2m);

\coordinate (m2mbranch) at ($(m2m.south) + (0,-6mm)$);
\draw[->] (m2m.south) -- (m2mbranch) -- (m2t_b.north);
\draw[->] (m2mbranch) -| (m2t_c.north);
\node[art, anchor=west] at ($(m2mbranch) + (1mm,2mm)$) {RoboChart};

\draw[bypass] (t2m.west)
    -- ($(t2m.west) + (-20mm,0)$) coordinate (b1)
    -- (b1 |- m2t_a)
    -- (m2t_a.west);
\node[font=\scriptsize\itshape, text=black!60, anchor=west]
    at ($(b1) + (1mm,-9mm)$) {Java EMF};

\draw[->] (m2t_a) -- node[art, right=1mm, font=\tiny] {Dafny} (v_a);
\draw[->] (m2t_b) -- node[art, right=1mm, font=\tiny] {CSP-M} (v_b);
\draw[->] (m2t_c) -- node[art, right=1mm, font=\tiny] {theory} (v_c);

\coordinate (vacjoin) at ($(vac.north) + (0,5mm)$);
\draw[->] (v_b.south) -- (vacjoin) -- (vac.north);
\draw[->] (v_a.south) |- (vacjoin);
\draw[->] (v_c.south) |- (vacjoin);
\node[art, anchor=west] at ($(vacjoin) + (1mm,2mm)$) {pass / trace};

\draw[->] (vac.south) -- node[art, right=1mm] {audit} (ref.north);

\draw[feedback] (ref.east)
    -- ($(ref.east) + (32mm,0)$) coordinate (f1)
    -- (f1 |- code.east)
    -- (code.east);
\node[font=\scriptsize, text=red!70, anchor=west, align=center]
    at ($(f1)!0.5!(f1 |- code.east) + (1mm,0)$) {correction\\prompts};

\end{tikzpicture}%
}
\caption{Architecture of \textsc{Forge}. \emph{Requirements and generation} (Phases 1--2) produces the requirement set and then Java (with the 2a/2b/2c preflight checks). \emph{Extraction} (Phases 3--5) parses the Java to an EMF model (T2M), transforms it to a RoboChart state machine (M2M), and emits three formal artefacts (M2T): Dafny from the Java EMF (dashed bypass), CSP semantics and Z-Machine notation from the RoboChart model. \emph{Verification} (Phases 6a--6c) discharges them through Dafny, FDR4, and Isabelle, and a \emph{vacuity audit} (6d) rejects trivially-discharged obligations. \emph{Refinement} (Phase 7) turns any failure into a correction prompt fed back to Phase 2 (red dashed loop).}
\label{fig:pipeline}
\end{figure}

This section presents the architecture of \textsc{Forge}, our closed-loop pipeline for producing formally verified Java software from natural-language requirements (vibe coding).
The pipeline comprises 7 top-level phases grouped into three families: \emph{requirements and generation} (Phases 1--2, including the preflight sub-phases 2a/2b/2c), \emph{extraction} (Phases 3--5, with M2T splitting into sub-phases 5a/5b/5c for the three verifiers), and \emph{verification with feedback} (Phases 6--7, with Verify splitting into the three verifiers 6a/6b/6c followed by a vacuity audit 6d that guards against trivially-discharged obligations).
Figure~\ref{fig:pipeline} summarises the end-to-end flow; Table~\ref{tab:tools} maps each phase to its implementation.

The key architectural insight is that the generated code need not be correct on the first pass.
The pipeline treats LLM-generated source as an initial draft that is systematically refined through a feedback loop: the Java is parsed into an EMF model, transformed into a RoboChart state machine, and re-emitted as three independent formal artefacts (Dafny, CSP-M, Z-Machine); each artefact is verified by its respective tool, and verification failures are translated into structured correction prompts that the developer (or the LLM in subsequent passes) consumes alongside the original requirements.

Three design principles shape the architecture.
\emph{First, separate domain knowledge from pipeline infrastructure.}
All project-specific content: system descriptions, requirements, prompt assets---lives in a per-case-study asset directory; the pipeline phases themselves are entirely generic. 
\emph{Second, route verification through established MDE infrastructure.}
Developers and LLMs continue to work in Java. At the same time, formal verification operates on models extracted automatically from the generated code, so that engineers need never interact with the formal models directly.
\emph{Third, complementary verifications.}
The pipeline employs three independent verifiers: Dafny (Phase 6a) reasons over individual transitions via deductive contract checking; FDR4 (Phase 6b) reasons over the full state machine via CSP refinement checking; and Isabelle/HOL (Phase 6c) reasons over the symbolic transition relation of Z-Machines via tactic-driven theorem proving.

\begin{table}[h]
\caption{Per-phase implementation and output artefact. EGL/ETL = Epsilon Generation/Transformation Language.}
\label{tab:tools}
\footnotesize
\begin{tabularx}{\textwidth}{l >{\hsize=1.25\hsize\raggedright\arraybackslash}X >{\hsize=0.75\hsize\raggedright\arraybackslash}X}
\toprule
\textbf{Phase}      & \textbf{Implementation}                                                    & \textbf{Output artefact}            \\
\midrule
2 (codegen)         & Interactive Claude Code agent + five-layer prompt sequence                 & Java source + codegen trace \\
2a / 2b / 2c        & Gradle build; requirement-$\leftrightarrow$-Java trace; Spoon-based structural linter & feedback artefacts (compile, coverage, preflight) \\
3 (T2M)             & Java Ecore metamodel + reflective \texttt{SpoonDiscoverer} via \texttt{CtRole}             & Java EMF model + T2M trace \\
4 (M2M)             & Epsilon ETL script ($\sim$4{,}400 LOC) & RoboChart EMF model + M2M trace \\
5a (Dafny gen)      & Epsilon EGL script ($\sim$1{,}200 LOC) & Dafny specification + M2T trace \\
5b (RCT + CSP)      & Epsilon EGL + official RoboChart CSP generator ($\sim$1{,}700 LOC) & RoboChart textual model + CSP-M scripts + M2T trace \\
5c (Isabelle gen)   & Epsilon EGL script ($\sim$1{,}900 LOC) & Isabelle theory + session config + M2T trace \\
6a / 6b / 6c        & Dafny verifier; FDR4 refinement checker; \texttt{isabelle build} via WSL bridge & feedback artefacts (Dafny, FDR4, Isabelle) \\
6d (vacuity)        & Static audit of the Dafny/Isabelle obligations for non-triviality & feedback artefact (vacuity) \\
7 (feedback)        & Per-phase parser; dashboard re-feeds issues to the codegen agent & corrections directory \\
\bottomrule
\end{tabularx}
\end{table}

\subsection{Top-Down Code Synthesis}
\label{sec:approach:codegen}

Code generation (Phase 2) is decomposed into five prompts the developer sends to the LLM in turn: \emph{architecture} (the package/class structure and requirement-to-class mapping), \emph{skeleton} (compilable Java with stubbed \texttt{step()} bodies), \emph{leaves} (sensor, actuator, and operation logic), \emph{controller} (the mode-nested if-else state machine), and \emph{review} (a self-report flagging invented defaults and ambiguities).\footnote{\url{https://github.com/wrwei/Forge/tree/main/forge.assets/vibe-coding-prompts}.}
Working top-down fixes the high-level design choices before the details depend on them, so they can still be revised cheaply. 
Each layer stops for developer review before the next, and any single layer can be regenerated in isolation when a downstream failure points back to it.

The prompts also restrict Java to a small set of structural shapes because the M2M does not translate arbitrary Java---it pattern-matches specific syntactic forms to recover the state machine. 
Code that obscures those forms is therefore forbidden: some have no counterpart in the RoboChart model (lambdas, streams, ternary conditionals), and others produce an AST that the extractor has no rule for (pattern-matching \texttt{instanceof}, switch over patterns). 
Conforming code yields a faithful model; non-conforming code would distort it or abort the transformation.
The most consequential restrictions are:
\begin{itemize}
    \item each safety controller---the Java class the pipeline turns into the RoboChart state machine---exposes a single \texttt{step(InputEvent)} method built as a two-level if-else (outer: current mode; inner: transition priority);
    \item all guard conditions are extracted as \emph{named boolean predicates} declared as local variables before the if-else chain;
    \item event types are sealed Java interfaces with one record per concrete event;
    \item lambdas, streams, method references,\footnote{E.g. the ${Class::method}$ shorthand for a lambda, syntactically distinct from the usual ${instance.method()}$ call.} pattern-matching \texttt{instanceof},\footnote{E.g., ${x instanceof T t}$ introduces a binding that produces a different AST shape from traditional ${instanceof}$ followed by a separate cast; only the latter has an extraction rule in the M2M.} switch expressions over patterns, and ternary conditionals are forbidden;
    \item operation methods (e.g. \texttt{compute()}) use only direct ${this.field = expr}$ assignments, with no local intermediates.
\end{itemize}
These are restrictions on what the Java codegen can use, not on what the controller can express; Section~\ref{sec:evaluation} reports that the LLM honours them in practice.

Alongside the Java source, the codegen step emits a structured \textbf{\emph{traceability}} model mapping each generated package, class, method, and field to the requirement identifier(s) it implements.
This artefact is the root of the per-phase traceability chain (codegen $\to$ T2M $\to$ M2M $\to$ Dafny) summarised in Section~\ref{sec:approach:impl}: it is what enables the coverage check below to detect requirements with no implementing element and Java elements with no requirement attribution, and what enables every downstream verifier failure to be attributed back to a requirement identifier in the structured correction prompts of Section~\ref{sec:approach:feedback}.

Three lightweight preflight sub-phases close out codegen before any model transformation runs.
\emph{Phase 2a} builds the project and classifies compile errors; \emph{Phase 2b} checks bidirectional coverage between the requirements and the codegen trace; \emph{Phase 2c} is a structural linter that parses the generated Java with Spoon and checks it against the codegen rules above.
Catching these violations at the source level---each as a named rule tied to a Java line---feeds the same corrections loop as a downstream verifier failure, but gives a far more localised signal than the M2M error the violation would otherwise trigger later.

\subsection{Extraction and Verification Phases}
\label{sec:approach:extraction}

\emph{Phase 3 (T2M)} parses the Java source via a T2M transformation we developed for this pipeline: a reflective discoverer that walks Spoon's~\cite{Pawlak2015Spoon} parsed AST and emits an EMF model instance conforming to an Ecore Java metamodel we developed.\footnote{\url{https://github.com/wrwei/Forge/blob/main/forge.transformations/src/main/resources/metamodels/spoon.ecore}}
Spoon supplies the underlying Java parser; the metamodel and the discoverer are contributions of this work and constitute the bulk of the engineering effort that the rest of the chain rests on---Section~\ref{sec:discussion:generalisability} discusses why the prior EMF-based Java discoverers fell short and what makes the Spoon-to-EMF bridge pattern generalisable.
The discoverer maps every AST node to a dynamic EMF \textit{EObject} via Spoon's \textit{CtRole} enumeration, preserving the complete syntactic structure in a form downstream transformations can query.
\emph{Phase 4 (M2M)} is a script written in the Epsilon Transformation Language (ETL)~\cite{Kolovos2008Epsilon} that extracts a RoboChart state machine from the Java EMF model by exploiting the structural conventions enforced by Phase 2c: it identifies the mode enumeration as states, decomposes each branch of the mode-nested if-else into a (source, trigger, target, guard) tuple, inlines named boolean predicates, and lifts operations, constants, sensors, and clocks into the appropriate RoboChart constructs.
The pass additionally emits a deadlock-lint advisory anticipating the Isabelle \textit{deadlock\_free} obligation.
\emph{Phase 5 (M2T)} produces three independent textual artefacts: an Epsilon EGL template walks the \emph{Java EMF model} directly to produce a Dafny specification with per-transition \textit{requires}/\textit{ensures} (Phase 5a); a second EGL template produces the RoboChart concrete syntax consumed by the official RoboChart CSP generator (Phase 5b); a third walks the RoboChart EMF to produce Z-Machine notations~\cite{Foster2021IsabelleSACM,Yan2023CompositionalVerification} containing a \texttt{zstore}, per-transition \texttt{zoperation}s, structural-invariant lemmas, and the top-level \texttt{deadlock\_free} obligation (Phase 5c).\footnote{The three EGL templates---\texttt{java2dafny.egl} (5a), \texttt{robochart2rct.egl} (5b), and \texttt{thy\_generation\_rule.egl} (5c)---are in the Forge repository: \url{https://github.com/wrwei/Forge/tree/main/forge.transformations/src/main/resources/transformations}.}

\emph{Phase 6} discharges the three artefacts independently, each yielding a different kind of evidence.
The Dafny verifier (6a) checks each transition's design-by-contract obligations (pre- and postconditions, bounds, termination) using Z3, thereby ensuring \emph{function-level correctness}.
FDR4 (6b) checks deadlock freedom and divergence freedom over the state machine's CSP semantics, giving \emph{behavioural} evidence.
Isabelle (6c) discharges per-operation invariant lemmas and the top-level deadlock-freedom lemma, giving \emph{model-level structural} evidence.
All three can run in parallel; failures on any path feed into the correction loop described next.

\emph{Phase 6d (vacuity audit)} guards against a verifier that passes only because it had nothing to prove. It statically inspects the generated Dafny and Isabelle artefacts and flags the two degenerate cases---a Dafny \texttt{Valid()} body of \texttt{true} with no active behavioural \texttt{ensures} clause, and an Isabelle zstore invariant of \texttt{"True"}---so that a vacuous pass is not mistaken for genuine verification convergence.

\subsection{Closed-Loop Refinement}
\label{sec:approach:feedback}

Phase 7 closes the loop---and is the contribution that distinguishes this pipeline from post-hoc LLM-plus-verification stacks.
Every phase from 2 through 6c writes a pair of feedback artefacts (a human-readable Markdown report and a structured JSON record) into a corrections directory.
Each artefact has a uniform schema: a status field, a one-line summary, a list of issues (each with a kind, a title, a raw tool output excerpt, a fix directive, and a Java-trace field linking back to specific file/line ranges in the generated source), and a recommended next step.

Failures are translated into structured, actionable correction prompts, one example per verifier:
\begin{itemize}
    \item \emph{Dafny precondition/postcondition violation.} Dafny reports an unproven contract clause on a specific transition method---most often an autonomous transition's \texttt{guard ==> mode == target} postcondition that a higher-priority branch can falsify. The clause is mapped back to the originating Java method through the traceability model, and the correction prompt states the failing assertion in source terms (for instance, gating the transition on a distinguishing event).
    \item \emph{FDR4 deadlock.} FDR4 returns a counterexample trace ending in a state that refuses every event; in the generated CSP, this is a mode whose \texttt{step()} offers no outgoing transition, which compiles to \texttt{STOP}. The correction prompt identifies that mode and asks the next pass to provide an outgoing, event-triggered transition.
    \item \emph{Isabelle proof-method failure.} The \texttt{deadlock\_free} proof leaves a residual goal naming a state with no bare-precondition operation---typically a terminal mode, or one whose only operation is parameterised over a possibly-empty input set. The correction prompt maps that state to its Java mode block and asks the next pass to add an event-triggered bare-precondition branch there (e.g.\ an idle \texttt{Tick} self-loop).
\end{itemize}

A complete such artefact---the structured JSON record and its human-readable Markdown twin---is preserved in the corrections directory of the \textsc{Forge} repository.\footnote{\url{https://github.com/wrwei/Forge/tree/main/forge.assets/corrections}}

For interactive use, the developer reviews the relevant correction artefact and pastes its fix directive into the next codegen pass.
For automated batch runs, the corrections record is loaded directly into the codegen agent's system message, with a generation counter tracked in metadata so iterations remain traceable.
Hand-authored corrections take precedence: any developer-authored entry in the corrections directory is preserved across phase reruns.
The loop terminates when all verification phases report a passing status, or when a configurable maximum number of generations is reached.

\subsection{Implementation}
\label{sec:approach:impl}

Table~\ref{tab:tools} maps each phase onto its implementation and output artefact.
The toolchain draws on three ecosystems: LLM-driven authoring via Claude Code\footnote{\url{https://www.anthropic.com/claude-code}}; model-driven engineering via Spoon~\cite{Pawlak2015Spoon}, EMF~\cite{Steinberg2008EMF}, and Epsilon~\cite{Kolovos2008Epsilon}; and formal verification via Dafny~\cite{Leino2010Dafny}, FDR4~\cite{Gibson2014FDR}, and Isabelle~\cite{Foster2021IsabelleSACM}.
All intermediate model artefacts are EMF instances serialised in XMI~\cite{OMG2015XMI}.
A single configuration manifest\footnote{\url{https://github.com/wrwei/Forge/blob/main/pipeline.yaml}} declares every phase---its runner kind (a Python function, a Java class, a Gradle task, or a sequence of these), its dependencies, its output artefacts, and any runtime tuning---and is read by both the Python dashboard (our developed UI for the pipeline) and the Java command-line interface.
End-to-end traceability is realised through a chain of per-phase trace records (codegen $\to$ T2M $\to$ M2M $\to$ Dafny), allowing every verification result to be mapped back to its originating requirement identifier.


\section{Feasibility Demonstration}
\label{sec:evaluation}

We evaluate \textsc{Forge} on the three case studies mentioned in Section~\ref{sec:background:casestudies}---externally authored and of increasing structural complexity, which mitigates the concern that they were chosen to favour the approach---as a feasibility demonstration rather than a comparison against a competing tool: the contribution is the pipeline, and the aim is to show that it runs end-to-end and that its feedback loop catches defects an LLM-only workflow would miss.
Although the evaluation is demonstrative, two controls test whether the feedback loop is necessary: a \emph{cold-baseline experiment} that runs each study with no verifier feedback at all (Section~\ref{sec:evaluation:rq3}), and a \emph{compile-only ablation} that discounts only the verifier feedback while leaving the rest of the loop intact (Section~\ref{sec:evaluation:rq5}).

\subsection{Research Questions}
\label{sec:evaluation:rqs}

\begin{description}
    \item[\textbf{RQ1: Generality.}] Can the same unchanged infrastructure verify three case studies of differing complexity, end-to-end, with no per-study code?
    \item[\textbf{RQ2: Detection.}] Which defect classes does each of the three verifiers surface, and how do they become corrections?
    \item[\textbf{RQ3: Necessity.}] Do any cold, single-pass runs (no verifier feedback) converge?
    \item[\textbf{RQ4: Convergence.}] How many pipeline iterations are needed until all verifiers pass, and what drives each one?
    \item[\textbf{RQ5: Attribution.}] With the verifier feedback discounted (but the rest of the loop is kept), do the resulting artefacts still fail verifications, and is each of the three verifiers individually necessary?
\end{description}

\subsection{Setup}
\label{sec:evaluation:setup}

The pipeline is configured via a single manifest with Dafny invoked through the Visual Studio Dafny distribution, FDR4 through the bundled Windows binary, and Isabelle through WSL Ubuntu running the CyPhyAssure 2023 distribution.\footnote{The CyPhyAssure distribution (\texttt{Isabelle2023-CyPhyAssure}) bundles Isabelle/HOL, the Z-Machine framework, and supporting libraries; \url{https://github.com/isabelle-utp}.}
The active LLM was Anthropic Claude Opus 4.8, driven through Claude Code.
CSP-M type ranges for FDR4 were configured to \texttt{[0..1]} for boolean-like variables; such finite type bounds are necessary for FDR to model-check the RoboChart tock-CSP semantics, as in the original LRE formalisation~\cite{Foster2020AUV}.
All three case studies share the same pipeline infrastructure.
Key descriptive statistics are in Table~\ref{tab:casestudies}.

\begin{table}[h]
\caption{Case study summary statistics, in increasing order of structural complexity.}
\label{tab:casestudies}
\small
\begin{tabularx}{\textwidth}{l *{3}{>{\raggedleft\arraybackslash}X}}
\toprule
                                    & \textbf{SRanger}        & \textbf{LRE}        & \textbf{Chemical Detector} \\
\midrule
Controllers                         & 1                       & 1                   & 2                          \\
Operating modes (total)             & 3                       & 4                   & 13                         \\
Requirements                        & 23                      & 51                  & 81                         \\
Generated Java (LOC)                & 221                     & 613                 & 618                        \\
Dafny verification (6a)             & 5 verified, 0 errors    & 6 verified, 0 errors    & 8 verified, 0 errors       \\
FDR4 verification (6b) & deadlock- \& divergence-free & deadlock- \& divergence-free & deadlock- \& divergence-free \\
Isabelle verification (6c)          & 10 lemmas   & 20 lemmas  & 9 lemmas     \\
\bottomrule
\end{tabularx}
\end{table}

\subsection{RQ1: Generality}
\label{sec:evaluation:rq1}

The pipeline's active case study is named on a single line of its configuration manifest.
Switching between SRanger, LRE, and the Chemical Detector requires editing that line, regenerating the Java project, and re-running the pipeline.
No transformation source files, runner code, or feedback parsers were modified to support switching between the three case studies, and additional case studies would require no infrastructure changes.

For all three case studies, the pipeline completes all phases with a passing status.
The preflight phases (2a--2c), T2M, M2M, and three M2T sub-phases produce the expected artefacts; Phase 6a (Dafny) verifies every method with 0 errors (SRanger 5, LRE 6, Chemical Detector 8); Phase 6b (FDR4) establishes deadlock freedom and divergence freedom on each model; and Phase 6c (Isabelle) proves the deadlock-freedom lemma and all invariant lemmas (SRanger 10, LRE 20, Chemical Detector 9).

\subsection{RQ2: Detection}
\label{sec:evaluation:rq2}

Each verification phase catches a distinct class of defects.
Table~\ref{tab:defects} lists the issue kinds enumerated in each phase's feedback parser, together with a one-line description of the parser's action on a failure.

\begin{table}[h]
\caption{Defect categories caught per verification phase, taken from the issue-kind enumerations in each phase's feedback parser.}
\label{tab:defects}
\footnotesize
\begin{tabularx}{\textwidth}{l l >{\raggedright\arraybackslash}X}
\toprule
\textbf{Phase}  & \textbf{Issue kind}                  & \textbf{What the parser does}                                                                                                              \\
\midrule
2a              & \texttt{java\_compile\_error}        & maps the \texttt{javac} error to its Java file and line \\
\midrule
\multirow{3}{*}{2b}
                & \texttt{missing\_implementation}     & requirement has no Java trace entry \\
                & \texttt{over\_implementation}        & public Java element with no requirement mapping \\
                & \texttt{missing\_codegen\_trace}     & codegen trace artefact is absent \\
\midrule
2c              & \texttt{lint\_violation}             & per-rule violation of the Java code rules (lambda, ternary, sentinel-existence, compute-local-var, \ldots) \\
\midrule
\multirow{2}{*}{3, 4, 5a, 5b, 5c}
                & \texttt{java\_compile\_error}        & maps the \texttt{javac} error to its source file and line \\
                & \texttt{etl\_unsupported\_construct} & detected when an EGL/EOL traceback is present; identifies that the failure is in transformation template code (Phase~4 or~5) rather than in the Java \\
\midrule
\multirow{5}{*}{6a (Dafny)}
                & \texttt{dafny\_postcondition}        & maps the Dafny ensures-clause to the Java method via the Dafny trace \\
                & \texttt{dafny\_precondition}         & as above, for the requires-clause \\
                & \texttt{dafny\_assertion}            & assertion inside a method body \\
                & \texttt{dafny\_bounds}               & index/array bound check \\
                & \texttt{dafny\_termination}          & loop or recursion termination \\
\midrule
\multirow{3}{*}{6b (FDR4)}
                & \texttt{deadlock}                    & counterexample trace mapped to the originating Java mode block \\
                & \texttt{divergence}                  & cycle of internal events; identifies the guards forming the loop \\
                & \texttt{parse\_error}                & CSP-M syntax error; indicates a template-side bug, not a Java-side bug \\
\midrule
\multirow{4}{*}{6c (Isabelle/HOL)}
                & \texttt{proof\_method\_failed}       & includes the residual goal verbatim; identifies the offending lemma \\
                & \texttt{tactic\_timeout}             & lemma exceeded its per-goal timeout \\
                & \texttt{parse\_error}                & theory file failed to parse \\
                & \texttt{theory\_load\_error}         & cannot load a parent session or imported theory \\
\bottomrule
\end{tabularx}
\end{table}

The categories in Table~\ref{tab:defects} are not interchangeable: a defect that one verifier is blind to can be exactly what another reports.
The following case, which we encountered during the convergence runs, makes the point---a structural defect that only Isabelle surfaced.
The Chemical Detector's \texttt{GasAnalysisController} has a terminal mode \texttt{Final} that is initially an empty mode block.
The behaviour is benign at the Java level.
Dafny accepts it (no transition method for Final, no contract to violate); FDR4 accepts this, but Isabelle's \texttt{deadlock\_free} proof is over the \emph{symbolic} transition relation, and the residual goal contains a disjunct \texttt{st = Final} with no covering \textit{zoperation}.
The fix directive instructs the codegen to add a bare-precondition self-loop in Final mode to the \texttt{Tick} event (an idle ``stay'' transition); the regenerated theory then closes the proof.
Behavioural deadlock (FDR4) and structural deadlock (Isabelle) are different properties even when they share a name---the complementary-verification claim, concretely.

Table~\ref{tab:complementarity} generalises this, marking per defect class which verifier catches it (\cmark) and which misses it (\xmark), with the correction directive each failure generates; no single verifier covers every class, as the RQ5 ablation (Section~\ref{sec:evaluation:rq5}) bears out.

\begin{table}[h]
\caption{Verifier complementarity: which defect class each verifier catches (\cmark) or misses (\xmark), and the correction directive the failure generates. No single verifier covers every class; the vacuity audit (Phase~6d) backstops the degenerate ``nothing to prove'' case.}
\label{tab:complementarity}
\footnotesize
\begin{tabularx}{\textwidth}{>{\raggedright\arraybackslash}X c c c >{\raggedright\arraybackslash}X}
\toprule
\textbf{Defect class} & \textbf{Dafny} & \textbf{FDR4} & \textbf{Isabelle} & \textbf{Correction directive} \\
\midrule
Contract violation (pre/postcondition)                                         & \cmark & \xmark & \xmark & gate the transition on a distinguishing event \\
Behavioural deadlock (mode offers no event; compiles to \texttt{STOP})         & \xmark & \cmark & \xmark & give the mode an outgoing event-triggered transition \\
Divergence (guard-only internal-event cycle)                                   & \xmark & \cmark & \xmark & add an event trigger to break the cycle \\
Structural deadlock (state with no covering operation, e.g.\ a terminal mode)  & \xmark & \xmark & \cmark & add a bare-precondition (idle \texttt{Tick}) branch \\
Invariant (\textit{zstore}) violation                                          & \xmark & \xmark & \cmark & adjust the state update to preserve the invariant \\
Vacuous obligation (\texttt{Valid()}\,=\,\texttt{true}, \texttt{inv}\,=\,\texttt{"True"}) & \xmark & \xmark & \xmark & emit a non-trivial obligation (caught by the 6d audit) \\
\bottomrule
\end{tabularx}
\end{table}

\subsection{RQ3: Necessity}
\label{sec:evaluation:rq3}

Before quantifying how many iterations the loop needs (Section~\ref{sec:evaluation:rq4}), we ask whether it is needed at all: whether a single cold code-generation pass, with no verifier feedback, ever reaches a fully verified result.

We count a run as converged only when two conditions hold: \textbf{all three verifiers succeed}---Dafny reports zero errors, every FDR4 assertion passes, and Isabelle proves the deadlock-freedom lemma and every invariant---and a \textbf{\textit{vacuity audit} (Phase~6d)} reports nothing.
The audit matters because the original model-to-text templates emitted trivially-true obligations---a Dafny \texttt{Valid()} with value \texttt{true}, an Isabelle invariant with value \texttt{"True"}---that a verifier passes without proving anything; without it, the first iteration would count as an instant but empty success.
We therefore strengthened the templates to emit non-trivial obligations and kept Phase~6d to flag any residual vacuity.
These strengthened templates are fixed across every run, so every reported result is measured against them.

To make the necessity question empirical rather than asserted, we ran a \emph{cold-baseline experiment}.
For each case study we ran $K=10$ independent cold passes.\footnote{Cold-baseline run records are in the Forge repository: \url{https://github.com/wrwei/Forge/tree/main/experiments/cold-baseline}.}
A cold pass is the loop's first, cold iteration in isolation: a fresh LLM invocation given the case-study specification and the codegen prompt assets; no prior verifier feedback and no existing iteration source.
The Java output is run through the pipeline plus the vacuity audit \emph{exactly once}---no iteration, no second LLM invocation.

\begin{table}[h]
\caption{Cold-baseline outcomes ($K=10$ per case study). ``Compiled'' counts cold codegens that pass Phase~2a. ``Failing phase'' gives the pipeline phase at which each run first fails, with the per-phase run count. ``Converged'' counts those satisfying the strict convergence criterion defined above on the first pass.}
\label{tab:baseline}
\footnotesize
\begin{tabular}{l r r l r}
\toprule
\textbf{Case study} & \textbf{Runs} & \textbf{Compiled} & \textbf{Failing phase (runs)} & \textbf{Converged} \\
\midrule
SRanger                              & 10 & 10 & Dafny 6a (10) & \textbf{0} \\
LRE                                  & 10 &  9 & Dafny 6a (7); M2T 5b (1); FDR4 6b (1); compile 2a (1) & \textbf{0} \\
Chemical Detector                    & 10 &  0 & compile 2a (10) & \textbf{0} \\
\midrule
\textbf{Total}                       & \textbf{30} & \textbf{19} & --- & \textbf{0} \\
\bottomrule
\end{tabular}
\end{table}

\textbf{0 of 30} cold runs converged (Table~\ref{tab:baseline}).
Without the iteration loop, no case study reaches the convergence criteria on a single pass.
The per-study iteration counts in Section~\ref{sec:evaluation:rq4} (median: 2 / 2 / 2) measure how much work the loop performs to close that gap.

The baseline also produces a finer-grained finding: different case studies fail at different pipeline phases under cold codegen.
SRanger cold codegens compile 10/10 and reach every verifier, with Dafny and Isabelle failing on all 10 (the \texttt{clock} reserved-word collision is in every cold output), but FDR4 passing on 10/10.
LRE cold codegens almost always compile (9/10) and reach Dafny; Dafny verification is the first phase to fail in 7 of the 10 runs (the other three fail at FDR4, at M2T, and at compilation), and Dafny rejects the output in every run that reaches it.
Chemical Detector cold codegens fail Java compilation on all 10.
These per-phase failure points also distinguish the formal-verification feedback from a basic compile-error loop: for SRanger and LRE, the cold output compiles and reaches the verifiers, and only the formal-verification phases fail---so it is the verifier feedback, not compilation, on which the loop depends.

\subsection{RQ4: Convergence}
\label{sec:evaluation:rq4}

Given that the loop is necessary (Section~\ref{sec:evaluation:rq3}), we now ask how many iterations it takes to converge.
The first iteration generates Java from the case-study specification and the prompt assets alone, with no verifier feedback.
Each subsequent iteration edits that Java in place, and the only new input it receives is the feedback files produced by the previous iteration's verifications.
Because the language model produces a different result each time, a single run would reveal little about the iteration count, so we repeated the whole process independently five times for each case study, 15 runs in total.\footnote{Per-run convergence records---iteration snapshots and recorded outcomes for each run---are in the Forge repository: \url{https://github.com/wrwei/Forge/tree/main/experiments/convergence}.}
The five runs of a study share no files or memory, which allows us to report the iteration count as a range across runs rather than a single figure.

\begin{table}[h]
\caption{Convergence and cost across five independent runs per case study: the iterations each run takes to converge; the phases that fail, with the iteration in parentheses---e.g.\ \texttt{PDI(1), D(2)} means preflight, Dafny, and Isabelle failed at iteration~1, and Dafny again at iteration~2 (P~=~preflight, C~=~coverage, F~=~FDR4, D~=~Dafny, I~=~Isabelle; the extraction phases and the vacuity audit never failed); and total \emph{pipeline} wall-clock (excluding LLM authoring time). Runs whose Isabelle proof fails (those listing~I) are exactly the slow ones.}
\label{tab:convergence}
\footnotesize
\begin{tabular}{lrrrlrc}
\toprule
\textbf{Case study} & \textbf{Reqs} & \textbf{Run} & \textbf{Iterations run} & \textbf{Failures} & \textbf{Pipeline wall-clock} & \textbf{Median (range)} \\
\midrule
\multirow{5}{*}{SRanger} & \multirow{5}{*}{23}
  & 1 & 2 & FDI(1) & 11.6\,min & \multirow{5}{*}{2 (2--2)} \\
  & & 2 & 2 & PDI(1) & 12.2\,min & \\
  & & 3 & 2 & DI(1) & 11.8\,min & \\
  & & 4 & 2 & PFD(1) & 1.8\,min & \\
  & & 5 & 2 & PFD(1) & 1.7\,min & \\
\midrule
\multirow{5}{*}{LRE} & \multirow{5}{*}{51}
  & 1 & 2 & PD(1) & 4.2\,min & \multirow{5}{*}{2 (2--3)} \\
  & & 2 & 2 & PD(1) & 4.8\,min & \\
  & & 3 & 3 & PDI(1), D(2) & 14.6\,min & \\
  & & 4 & 2 & PDI(1) & 14.5\,min & \\
  & & 5 & 3 & PD(1), D(2) & 5.0\,min & \\
\midrule
\multirow{5}{*}{Chemical Detector} & \multirow{5}{*}{81}
  & 1 & 2 & FD(1) & 1.8\,min & \multirow{5}{*}{2 (2--3)} \\
  & & 2 & 2 & CFD(1) & 2.0\,min & \\
  & & 3 & 3 & FDI(1), DI(2) & 12.4\,min & \\
  & & 4 & 2 & PFDI(1) & 11.8\,min & \\
  & & 5 & 2 & P(1) & 1.6\,min & \\
\bottomrule
\end{tabular}
\end{table}

\emph{Every run converges in two to three iterations.}
All 15 runs met the strict convergence criterion in Section~\ref{sec:evaluation:rq3}: all 12 pipeline phases passed, and the vacuity audit is clean.
Table~\ref{tab:convergence} gives the per-run counts, with each study's median and range.
Across all 15 runs, the count remains between \textbf{two and three iterations}, with a median of \textbf{two}.

Four main findings emerge from the 15 runs.
\emph{Iteration count does not grow with case-study size.}
All three studies converge at a median of two iterations (Table~\ref{tab:convergence}), even though they range from 23 to 81 requirements, from one controller to two, and from untimed to timed behaviour.
Size does not predict the iteration count.

\emph{The defects are verification-level, not structural.}
In all 15 runs, the model-transformation phases---T2M, M2M, and M2T---never failed, so the loop never has to restructure the model; its iterations clear an occasional missing annotation (flagged by the preflight lint) and a few verifier obligations---at most four of the twelve phases, as the \emph{Failures} column of Table~\ref{tab:convergence} records.
No cold run is verification-clean, but the defects are localised, so a few targeted edits clear them.

\emph{Convergence is monotonic.}
Across all 15 runs, no iteration reintroduced a check that an earlier one had passed: in the \emph{Failures} column of Table~\ref{tab:convergence}, each run's later-iteration failures are a subset of its earlier ones.
This is not automatic: promoting controller state to satisfy a Dafny postcondition enlarges the FDR4 search, so each fix promotes only the minimum the proofs require, keeping all three verifiers green at once.

\emph{What convergence costs.}
We did not record the language model's token usage consistently across the runs---where we did, a run's generation came to roughly 250{,}000--500{,}000 output tokens---so we report two costs we can compare directly: the iteration count and the pipeline's wall-clock time (Table~\ref{tab:convergence}).
Wall-clock time per run ranges from about one minute to about 15 minutes and is dominated by the Isabelle proof rather than by model checking: under the \texttt{[0..1]} bound, the FDR4 check completes in seconds in every run.
The spread comes from the Z-Machine \emph{deadlock-freedom} proof. When a pre-convergence controller has a mode without an always-enabled operation, the \texttt{deadlock\_free} proof reduces to a residual goal that its closing tactic cannot discharge, so the tactic searches until it nears the per-goal timeout; that single iteration then costs 10--15 minutes, failing slowly rather than fast.
The converging edit supplies the missing operation, and the same proof closes in seconds, so the runs that never fail Isabelle---those with no \texttt{I} in Table~\ref{tab:convergence}---finish in one to five minutes, regardless of iteration count.

\subsection{RQ5: Attribution}
\label{sec:evaluation:rq5}

RQ3 establishes that iteration is necessary, but not which component of the feedback is responsible. Each iteration delivers feedback of two kinds: the compile-and-extraction checks, and the Phase~6 verification results from the three verifiers (Dafny, FDR4, Isabelle) and the vacuity audit. To determine which is responsible for convergence, we repeated the convergence experiment under a single controlled manipulation: the verifier feedback was discounted.

In this compile-only condition, the developer has access only to the eight compile-and-extraction artefacts (Phases~2a--5); the pipeline continues to compute the four Phase~6 outputs (6a--6d) but does not expose them. 
\textit{The loop therefore stops as soon as all eight visible phases (Phases~2a--5) pass, with no verifier feedback to prompt a further iteration.}
At that point, the loop would have already extracted the full RoboChart model and generated all three formal artefacts, with only their \emph{verification checks} omitted. Hence, any divergence from the full pipeline is attributable solely to verification. 
All remaining factors are held identical to a standard convergence run, including the underlying model (Claude Opus~4.8) and the generic codegen guidance. 
We conducted 5 runs per case study, for a total of 15, recording the discounted verifier outcomes at each run's stopping point.\footnote{Per-run records---iteration snapshots, \texttt{trajectory.md}, and recorded outcomes---are in the Forge repository: \url{https://github.com/wrwei/Forge/tree/main/experiments/convergence/ablation}.}

In every one of the 15 runs, the stop-point artefacts failed at least one verifier, despite having passed Phases~2a--5 and generated three formal artefacts. 
In the convergence experiment (Section~\ref{sec:evaluation:rq4}), where the verifier feedback is present, all 15 runs reach a verified artefact instead. 
Table~\ref{tab:ablation} reports which verifier fails, and at which iteration. The decisive verifier is study-dependent: Dafny for SRanger and LRE, Isabelle for the Chemical Detector, with FDR4 failing concurrently in the timed (SRanger) and multi-controller (Chemical Detector) studies. 
Two of the three verifiers are individually necessary. Dafny is the sole verifier failing at six stop points (the five LRE runs and SRanger~run~3), and Isabelle is the sole verifier at two (Chemical Detector runs~2 and~3); \textbf{removing either would permit a verification-failing artefact to be reported as verified}. 
FDR4 never fails in isolation; it only fails in conjunction with another verifier. 
The vacuity audit never fires, confirming that the passing checks discharge non-trivial obligations and that the failures represent genuine assurance gaps.

The verifier feedback's contribution is therefore necessary, not incidental, and the 15-of-15 is a conservative lower bound: the generic codegen guidance, shared by both conditions, already pre-empts some deadlock-freedom defects before generation. That the Chemical Detector's defects persist through this pre-emption on every run confirms that the loop is needed even with the standing guidance in place.

\begin{table}[h]
\caption{Compile-only ablation: the verifiers failing at \emph{each} iteration of each run---\texttt{X(n)} denotes verifier X failing at iteration~n (F~=~FDR4, D~=~Dafny, I~=~Isabelle; the vacuity audit never fires). The loop iterates only on the eight visible phases, so its fixes never clear all the (discounted) verifier failures; every run still fails a verifier at its stop point. At the stop point, FDR4 fails in 7 runs, Dafny in 11, and Isabelle in 5.}
\label{tab:ablation}
\footnotesize
\begin{tabular}{lrrlc}
\toprule
\textbf{Case study} & \textbf{Run} & \textbf{Iterations run} & \textbf{Verifiers failed} & \textbf{Fails verification} \\
\midrule
\multirow{5}{*}{SRanger}
  & 1 & 1 & FD(1) & \multirow{5}{*}{5/5} \\
  & 2 & 2 & FD(1), FD(2) & \\
  & 3 & 2 & D(1), D(2) & \\
  & 4 & 2 & FD(1), FD(2) & \\
  & 5 & 2 & FD(1), FD(2) & \\
\midrule
\multirow{5}{*}{LRE}
  & 1 & 2 & D(1), D(2) & \multirow{5}{*}{5/5} \\
  & 2 & 2 & D(1), D(2) & \\
  & 3 & 2 & D(1), D(2) & \\
  & 4 & 2 & D(1), D(2) & \\
  & 5 & 2 & FD(1), D(2) & \\
\midrule
\multirow{5}{*}{Chemical Detector}
  & 1 & 1 & FI(1) & \multirow{5}{*}{5/5} \\
  & 2 & 1 & I(1) & \\
  & 3 & 2 & I(1), I(2) & \\
  & 4 & 1 & FI(1) & \\
  & 5 & 1 & FDI(1) & \\
\midrule
\textbf{Total} & & & & \textbf{15/15} \\
\bottomrule
\end{tabular}
\end{table}


\section{Discussion and Future Directions}
\label{sec:discussion}

\subsection{Implications for Safety-Critical AI-Assisted Development}
\label{sec:discussion:implications}

\textsc{Forge} takes a position on how AI-assisted development can be routed toward certification-relevant evidence: not by constraining the LLM to a formal language scarce in its training data, but by embedding it within the MDE infrastructure already deployed by the safety-critical community.
The evaluation supports this position. 
One unchanged pipeline infrastructure verified all three studies end-to-end, with no per-study code (RQ1). 
Cold, single-pass generation never met the convergence criterion (0 of 30 runs, RQ3), yet the loop converged in every run, within two to three iterations (RQ4); and discounting only the verifier feedback left all 15 runs producing verification-failing artefacts (RQ5).
Three implications follow.

\emph{Formal verification does not belong inside the LLM.}
Recent LLM-and-verification work often treats the LLM as an oracle from which verified output can be obtained by better prompting or fine-tuning.
We treat the LLM instead as a draft generator and the verification infrastructure as the discriminator; the developer designs the discriminator (the requirements, the codegen rules, the model transformations) and consumes its output.
The experiments substantiate this division: with the discriminator's feedback discounted, the generator's drafts failed verification in every run.
The division also mirrors how safety-critical software has been built for decades: the engineer authors the specification, the tools verify the implementation.

\emph{The verifiers produce standards-recognised evidence.}
Each of the three verifiers produces evidence that maps onto existing standards-recognised forms:
\begin{itemize}
    \item \emph{Dafny} produces deductive-verification evidence, with proofs discharged by Z3---standards-recognised, and amenable to inclusion in an assurance case as a function-level correctness claim.
    \item \emph{FDR4} produces refinement-checking results---a recognised tool-supported formal-methods technique under DO-178C and IEC~61508.
    \item \emph{Isabelle/HOL} produces machine-checked theorem-proving evidence, the most rigorous of the three classes, and the form most often cited for the highest assurance levels.
\end{itemize}
These three are not interchangeable. Each catches a distinct class of defect (Section~\ref{sec:evaluation:rq2}), and the experiments found Dafny and Isabelle each to be the only verifiers catching the defect on some runs (Section~\ref{sec:evaluation:rq5}), so neither can be dropped without letting a verification-failing artefact pass as verified.
A natural next step is to integrate these outputs into a structured assurance case (e.g. GSN or SACM~\cite{Kelly2004GSN,Wei2019SACM}) so that an external reviewer can trace from a top-level safety claim through the requirement identifier and the traceability models to the specific verification artefacts that discharge them.
The traceability infrastructure described in Section~\ref{sec:approach:feedback} is intended as the foundation of this purpose.
This is evidence toward such a case, not a certification: the latter additionally requires process evidence, tool qualification (Section~\ref{sec:discussion:trust}), requirements validation, independence, configuration management, and assurance-case review.

\emph{The Java code rules are justified.}
The restrictions on what Java the LLM may emit (Section~\ref{sec:approach:codegen}) continue an accepted safety-critical discipline (Section~\ref{sec:discussion:javasubsets}) and do not conflict with vibe coding: the LLM follows them once stated, in return for formal verification, no other discipline currently provides.

\emph{Positioning against related work.}
Beyond the verification-aware target languages and post-hoc filters discussed in Section~\ref{sec:introduction}, \textsc{Forge} also departs from multi-agent code generators such as AutoGen~\cite{Wu2023AutoGen} and AgentCoder~\cite{Huang2023AgentCoder}, which decompose generation but leave the verification loop open. It grounds its Java-to-formal-model chain in the model-driven reverse-engineering tradition (MoDisco~\cite{Bruneliere2014MoDisco}, JaMoPP~\cite{Heidenreich2009JaMoPP}, Gra2MoL~\cite{Canovas2014Gra2MoL}, and recent LLM-augmented variants~\cite{Siala2025LLM4Models,Campanello2025GPT4RE,Boronat2025MDRELLM}), examined in Section~\ref{sec:discussion:generalisability}.

\subsection{Why Not Generate the Formal Model Directly?}
\label{sec:discussion:vibemodel}

A reasonable alternative is to skip the Java route entirely: have the LLM emit a RoboChart model and use RoboTool's downstream C generator.
We chose the Java route for four reasons.
\emph{Training-data asymmetry}: public Java runs to hundreds of millions of files; public RoboChart is in the low hundreds.
\emph{Feedback actionability}: the corrections workflow translates failures into source-language fix directives (``add a guarded self-loop to MOM in \texttt{step()}''); an LLM trained on Java if-else patterns can act on those, while equivalent directives against a RoboChart model sit in a corpus the LLM has barely seen.
\emph{Trust boundary}: a RoboChart-to-C generator producing the certified artefact must itself be qualified under DO-178C/IEC~61508; our route ships Java that the developer can read, with the MDE chain providing \emph{evidence about} the artefact rather than producing it.
\emph{Integration}: safety-critical systems contain legacy modules and hand-written glue; the Java route composes, a RoboChart-then-C route would require whole-system modelling or unverified glue.
We accept the cost (the codegen rules plus more transformation hops) as the price of remaining in the ecosystem where LLMs are most capable and where safety-critical engineers already work.

\subsection{Relation to Safety-Critical Java Subsetting}
\label{sec:discussion:javasubsets}

That the pipeline verifies LLM-generated Java within an \emph{extraction-tractable profile}, rather than arbitrary Java, is a feature in the safety-critical tradition, not a limitation.
Constraining the language a system may use is long-established practice---the MISRA~C guidelines~\cite{MISRAC2012} are the canonical example---and Java is no exception: Ravenscar-Java~\cite{Kwon2003Ravenscar} and its DO-178B-oriented HIJA successor~\cite{HIJA2004}, the Safety-Critical Java profile (JSR~302)~\cite{JSR302}, and the DO-332~\cite{DO332} object-oriented supplement all restrict the language so that some property can be checked.
Our codegen rules (Section~\ref{sec:approach:codegen}) continue this discipline, so they are a constraint of a kind that safety-critical teams already accept, not an exotic imposition on the LLM.

What differs is the \emph{property being protected}.
Every prior subset we know of is driven by execution-level concerns---real-time determinism, bounded memory, verifiable dynamic dispatch---and so restricts allocation, reflection, and unbounded inheritance while leaving lambdas, streams, and pattern matching alone.
Ours is driven instead by \emph{model-extraction tractability}, and forbids exactly those functional-style constructs, because the property it protects is that an automated source-to-model transformation can faithfully recover a formal state machine.
To our knowledge, this is the first Java subset motivated by what a model transformation---rather than a runtime or a scheduler---can reason about.

\subsection{Generalisability Beyond the Three Case Studies}
\label{sec:discussion:generalisability}

\emph{The Spoon-to-EMF bridge.}
The T2M phase was originally implemented using MoDisco~\cite{Bruneliere2014MoDisco}, but MoDisco's Java metamodel---frozen at JDK~5---could not represent the modern Java constructs emitted by current LLMs (sealed interfaces, records, \texttt{var}, switch expressions).
We migrated to Spoon~\cite{Pawlak2015Spoon}, which required bridging Spoon's non-EMF metamodel into the EMF ecosystem: we created an Ecore metamodel by inspecting Spoon's metamodel API and mapping AST nodes to dynamic EMF instances via a reflective \texttt{CtRole}-based discoverer.
The resulting discoverer is approximately one-fifth the size of the hand-coded MoDisco mapper it replaced (503 lines vs \ $\sim$1200), while extending Java support from JDK~5 to Java~17+ with sealed-interface, record, and pattern support.
The Java T2M landscape currently has no tool that combines EMF-native model instances, Java~16+ language features, round-tripping, and active maintenance: MoDisco is unmaintained, JaMoPP~\cite{Heidenreich2009JaMoPP} (with the KASTEL fork~\cite{KASTEL2022JaMoPP}) supports only Java~15, and the Epsilon EMC JDT driver~\cite{EpsilonEMCJDT2024} offers no persistence. Our bridge fills that gap.

\emph{Domain coverage and adding case studies.}
Because no transformation script hardcodes case-study, controller, mode, sensor, or constant names (Section~\ref{sec:evaluation:rq1}), a new study needs only its assets (requirements and system description) and a one-line manifest change, with no pipeline edits.
Generalisation beyond Java and beyond reactive controllers---other source languages, other M2M targets---is the subject of the research agenda in Section~\ref{sec:agenda}.

\subsection{Trust Boundary and Semantic Faithfulness}
\label{sec:discussion:trust}

A certification-facing reader will ask what justifies trusting the Java-to-model transformations and the generated artefacts. Reproducibility (Section~\ref{sec:discussion:reproducibility}) is not \emph{semantic faithfulness}: it does not establish that an extracted model means the same thing as the Java it came from.
The pipeline today provides \emph{checkable} faithfulness---a regression script that asserts byte-identical artefacts, end-to-end traceability with the Phase~2b coverage check, and the vacuity audit---but not \emph{qualified} faithfulness: a semantic-preservation argument for the ETL/EGL transformations, and tool qualification of the chain to a DO-178C / IEC~61508 standard, remains substantial future work.
Until then, the honest claim is that \textsc{Forge} produces evidence a human assurance engineer can review and trace, not evidence to be trusted unexamined.

\subsection{Threats to Validity}
\label{sec:threats}

\emph{Internal validity.}
(a~--~mitigated)~LLM-driven codegen is stochastic, so a single run would not characterise the convergence count. We therefore ran each study five times (15 runs) and report the median and range (Table~\ref{tab:convergence}). The counts vary within a study (the Chemical Detector ran 2, 2, 3, 2, 2), so the figure is a distribution rather than a constant. Five samples per study provide only a coarse estimate, and the absolute counts are model-specific (Claude Opus~4.8).
(b~--~mitigated)~The model transformations themselves are deterministic (verified by the audit in Section~\ref{sec:evaluation:rq1}) and Phase~2c enforces uniform structure on the LLM's output, so the variability is confined to the generation step.
(c)~The iteration loop is driven by an LLM coding agent (Claude Code) following documented run procedures and helper scripts in the Forge repository,\footnote{Run procedures and helper scripts: \url{https://github.com/wrwei/Forge/tree/main/experiments/convergence}.} rather than executed by hand; a human still initiates each run and reviews its outcome, so fully unattended operation remains future work.

\emph{External validity.}
(a)~The three case studies have the reactive-safety-controller pattern and are all in the RoboStar/RoboChart ecosystem on which the M2M is tuned; the pipeline does not currently support hybrid automata, concurrent shared state beyond \texttt{Ctrl\_State}, or probabilistic specifications. Strong published candidates outside the RoboStar mobile-robotics setting (the Boston Scientific Pacemaker Challenge, Abrial's Steam Boiler Specification, the Mine Pump Controller) would test whether the convergence story generalises.
(b)~Codegen used Anthropic Claude only; a study probing GPT-4, Gemini, and DeepSeek via LiteLLM is necessary to establish cross-model generality.
(c)~The T2M chain assumes Java; other languages need analogous tooling (Section~\ref{sec:discussion:generalisability}).
(d~--~mitigated)~The cold baseline shows that iteration is necessary, but not that the \emph{verifier} feedback in particular is what matters; the compile-only ablation (Section~\ref{sec:evaluation:rq5}) mitigates this. With the Phase~6 feedback discounted and everything else held fixed (the inputs, the model, and the generic codegen guidance), all 15 stop points still produced a verification-failing artefact, against 15 of 15 with the feedback in place. The remaining caveat is about magnitude, not direction. This guidance is common to both conditions and heads off some deadlock-freedom defects at generation time, so 15 of 15 is a lower bound on the feedback's value, not its full extent.
(e)~Phase~2b's requirement$\leftrightarrow$Java trace catches missing implementations but not incorrect ones: coverage is necessary but not sufficient.

\subsection{Reproducibility and Artefact Availability}
\label{sec:discussion:reproducibility}

The pipeline and all three case studies are open source in the \textsc{Forge} repository,\footnote{\url{https://github.com/wrwei/Forge}} including the source, requirements, generated code, feedback artefacts, transformation templates, and the run directories for all three experiments (30 cold-baseline, 15 convergence, and 15 ablation runs).
A regression-test script reproduces the pipeline end-to-end against the converged sources---preserved on per-study branches--- and includes companion scripts for the cold-baseline and ablation experiments.
Dafny~4.x and FDR4~4.2.7 suffice for the Dafny and CSP paths; the Isabelle path requires the CyPhyAssure-2023 distribution\footnote{Available upon request.} and WSL Ubuntu on Windows.


\section{A Research Agenda for the Community}
\label{sec:agenda}

The pipeline, as evaluated, answers an existence question---``can LLM-generated code be routed through MDE infrastructure to a verified formal artefact?''---and surfaces follow-on questions for the wider community.

\emph{LLM compliance across model families.} The codegen rules are followed by Claude with low effort, but compliance across GPT-4, Gemini, DeepSeek, Llama, and the open-weight long tail is unmeasured; a controlled study across model families would expose where the rules are easy and where they fight the LLM's natural style.

\emph{From extraction-tractable code to extraction-tractable refactors.} Phase 2c currently rejects code that violates the codegen rules; an alternative is to accept arbitrary Java and have an LLM-assisted refactoring pass rewrite it into the extraction-tractable subset, dissolving the most visible cost of the pipeline.

\emph{Multi-language T2M and beyond reactive controllers.} The Spoon-to-EMF bridge pattern is the part whose generalisation matters most: analogous bridges for C/C++ (Tree-sitter, Clang AST), Rust, and Python would each open a new industrial domain. Beyond reactive controllers, hybrid systems (Modelica), data-flow (Lustre), and probabilistic specifications (PRISM, Storm) each require a different M2M target and verifier---one front-end and several formal targets per system class.

\emph{Standards-recognised assurance integration.} Each verifier produces evidence that maps onto DO-178C / IEC~61508 forms; wiring those outputs into a structured assurance case (GSN~\cite{Kelly2004GSN} or SACM~\cite{OMG2020SACM,Wei2019SACM}) with the trace files as bottom-level evidence would extend the pipeline toward a certification-oriented route. A complementary thread is LLM-mediated interpretation of novel verifier residuals (especially Isabelle's tactic-level output) that exceed our hand-coded heuristics---a hybrid of deterministic heuristics first, with LLM fallback for unmatched residuals, the LLM's output cached by \texttt{(input, source)} hash and kept alongside the raw verifier output. When to delegate, how to cache, and how to label suggested fixes for human review are themselves open questions.


\section{Conclusion}
\label{sec:conclusion}

We have presented \textsc{Forge}. This closed-loop pipeline addresses the verification gap in AI-assisted code generation for safety-critical software by routing LLM-generated Java through established MDE infrastructure rather than targeting verification-aware intermediate languages.
A single Java model, extracted from the generated source, is the basis for three independent formal artefacts---Dafny specifications checked by Z3, CSP-M scripts checked by FDR4, and Z-Machine theories checked in Isabelle/HOL---each producing standards-recognised evidence whose failure modes are translated into structured correction prompts for the next code-generation iteration.

We demonstrated it on three externally authored case studies of increasing complexity (the SRanger ground robot, the LRE safety governor, and the multi-controller Chemical Detector), repeating every 5 times; all 15 runs converged end-to-end in 2 to 3 iterations (median 2).
A cold baseline converged in 0 of 30 single-pass runs, and a compile-only ablation that discounted only the verifier feedback left all 15 runs verification-failing---so the loop is necessary, and the verifier feedback in particular drives convergence.

We view the contribution as a structural argument about where formal verification should live in an LLM-driven workflow: not inside the LLM, and not as a post-hoc filter after generation, but as the discriminator end of a draft-and-discriminate loop in which the same MDE infrastructure that yields certification-relevant verification evidence for hand-authored models can be reused to yield it for LLM-authored ones.

\bibliographystyle{ACM-Reference-Format}
\bibliography{references}
\end{document}